\documentclass[aps,prd,twocolumn,amsmath,amssymb,groupedadress]{revtex4-1}

\usepackage[utf8]{inputenc}
\usepackage[dvips]{graphicx}
\usepackage{psfrag}
\usepackage{color}

\newcommand{\dd}{\mathrm{d}}

\usepackage{xspace}

\usepackage{hyperref}

\begin{document}

% Head of the paper %%%%%%%%%%%%%%%%%%%%%%%%%%%%%%%%%%%%%%%%%%%%%

\title{Backreaction in Growing Neutrino Quintessence}

\author{Florian Führer}
\email[]{fuhrer@thphys.uni-heidelberg.de}
\author{Christof Wetterich}

\affiliation{Institut für Theoretische Physik,\\ Universität
Heidelberg, Philosophenweg 16, D--69120 Heidelberg, Germany}

\date{\today}

\begin{abstract}
We investigate the cosmological effects of neutrino lumps in Growing Neutrino
Quintessence. The strongly non-linear effects are resolved by means of
numerical N-body simulations which include relativistic particles,
non-linear scalar field equations and backreaction effects. For
the investigated models with a constant coupling between the scalar field and the neutrinos the backreaction effects are so
strong that a realistic cosmology is hard to realize. This points towards the necessity of a field dependent coupling in Growing
Neutrino Quintessence. In this case realistic models of dynamical Dark Energy
exist which are testable by the observation or non-observation of
large neutrino lumps.
\end{abstract}


\maketitle

%%%%%%%%%%%%%%%%%%%%%%%%%%%%%%%%%%%
%%%%%%%%%%%%%%%%%%%%%%%%%%%%%%%%%%%
%\listoftodos

\section{Introduction}
\label{sec:introduction}
The origin of the observed accelerated expansion of the universe is still
unknown \cite{Perlmutter99, Riess98}. It is usually accounted for by a Dark Energy (DE) component. The simplest possibility
consistent with observations is a cosmological constant $\Lambda$, but a lot of alternatives have been proposed
\cite{Copeland06}. Prime candidates are dynamical Dark Energy
models mediated by a scalar field or modified gravity - the latter being often equivalent to the
former \cite{Wetterich14b}. For many alternatives the cosmological constant
problem \cite{Weinberg89,Carroll01} of explaining the small value of
$\Lambda$ persists, however. Also the explanation of why DE becomes
important in the present cosmological epoch is often not more convincing than
for a cosmological constant.

Growing Neutrino Quintessence (GNQ) \cite{Amendola08, Wetterich07} offers here
some advantages. As a quintessence model
\cite{Wetterich88,Ratra88} the late time acceleration is driven by a scalar
field $\varphi$ (the cosmon), employing a mechanism similar to Inflation. It is
possible to unify the late and early time acceleration into a single picture
\cite{Wetterich13,Wetterich13b,Wetterich14} so that the same field is
responsible for DE and Inflation. As an overall description within quantum
gravity crossover cosmology \cite{Wetterich14c} GNQ also addresses the
cosmological constant problem.

GNQ is able to explain the smallness of the DE
component, since the dynamical DE density decays
during the cosmic history, just as the other energy densities in the universe.
The DE density being small is then just a matter of time - it is
small because the universe is old. In contrast to simpler Quintessence models
GNQ solves the Why-Now-Problem. No fine tuning of the self interaction potential
is needed for this purpose. A coupling between the cosmon and the neutrinos
provides a mechanism for stopping the evolution of the cosmon field as soon as the neutrinos become
non-relativistic. The phenomenology of a very slowly evolving scalar field
resembles a cosmological constant. The transition from relativistic to non
relativistic neutrinos acts as a trigger for the DE domination. For neutrino
masses allowed by observations this transition happens in the ``recent'' past,
explaining why DE has become important now.

Despite a background evolution similar to the $\Lambda$CDM model for
redshift $z\lesssim 5$, GNQ has a phenomenology which is distinct from other
models. It predicts a time varying neutrino mass and the formation of neutrino
lumps, which might be detectable through there gravitational potentials \cite{Mota08}. The formation
of lumps is a consequence of the large coupling between neutrinos and the
cosmon, which is required for the stopping mechanism. The resulting additional
attraction between neutrinos is about $10^3$ times stronger than the
gravitational attraction. It can have a natural explanation in a particle
physics framework \cite{Wetterich07}.

While the strong coupling on the one hand offers with the lumps a clear and
distinct way of testing the model. On the other hand, it renders the model
technically difficult to study. In GNQ perturbations in the neutrino density become non-linear already at $z\approx 1-2$ on very large scales \cite{Mota08}.
This has lead to the development of a comprehensive N-body simulation
\cite{Ayaita12a,Ayaita14} to follow the formation of the neutrino lumps. The
simulation is different from the usual CDM-only simulations: In order to include backreaction 
effects, induced by the highly non-linear nature of the lumps \cite{Pettorino10}, the background is
solved simultaneously with the perturbations.
Additionally, neutrinos becoming relativistic during the formation of lumps is
captured by the simulation. A similar framework for relativistic N-body
simulation with focus on the metric perturbations was explored recently in
\cite{Adamek13}. With our simulation it was possible to draw a
consistent picture of neutrino structures within GNQ. For stable lumps the main
characteristic features can be understood within an approximation in terms of a
non-relativistic fluid of neutrino lumps \cite{Ayaita12b}.

In this work we investigate if GNQ can provide a realistic expansion history.
Therefore we study the equation of state and the energy density of the cosmon
for different model parameters.
We aim at finding model parameters for which the backreaction effect remains
compatible with an accelerated expansion with $\Omega_{DE}\approx0.7$.
At the same time the accelerated expansion of the universe must start early enough to be consistent with observations.

A time dependent neutrino mass related to a scalar Dark
Energy field concerns a wider setting than GNQ. Mass varying neutrino
scenarios (MaVaNs) have been studied earlier in \cite{Fardon03} and share common
features with GNQ as the instability of neutrino perturbations
\cite{Afshordi05,Bjaelde07,Brouzakis07}.

This work is organized as follows. We start with a brief review of GNQ in
section \ref{sec:gnq}. In section \ref{sec:backreact} we discuss the
formation of lumps and their backreaction on the cosmological expansion.
In section \ref{sec:N-Body} we describe our simulation, which we use to perform
a parameter scan. Results are presented in section \ref{sec:Results}. Finally,
we conclude in section \ref{sec:conclusion}.
%%%%%%%%%%%%%%%%%%%%%%%%%%%%%%%%%%%
%%%%%%%%%%%%%%%%%%%%%%%%%%%%%%%%%%%

\section{Growing Neutrino Quintessence}
\label{sec:gnq}
\subsection{Basic Concepts}
\label{sec:basic}
In this section we briefly describe GNQ. The ingredients of GNQ are a scalar
field $\varphi$ (the cosmon) and neutrinos. The neutrino mass
depends on the value of $\varphi$, thereby coupling the cosmon and the neutrinos.
The cosmon itself is described by the standard Lagrangian of a scalar field
which takes, using the metric signature $(-,+,+,+)$ and setting the reduced
Planck Mass to unity, $8 \pi G =1$, the form:
\begin{align}
-\mathcal{L}_{\varphi}=\frac{1}{2}\partial_{\mu}\varphi\partial^{\mu}\varphi+V(\varphi).
\end{align}
We choose an exponential potential $V(\varphi)\propto e ^{-\alpha \varphi}$.
As long as the neutrino mass can be neglected the exponential potential leads to
scaling solutions of the cosmon field. The background energy density of the cosmon
becomes independent of the initial conditions and mimics matter (radiation)
during matter (radiation) domination \cite{Wetterich88}, where the energy
density of the cosmon is a constant fraction of the total energy density
$\Omega_{\varphi}=3\frac{1+w}{\alpha^2}$. Here $w$ is the equation of state of
the dominating species. Constraints on early dark energy (EDE) require
$\alpha\gtrsim 10$ \cite{Doran07,Reichardt12,Pettorino13,PlanckXIV15}, where we use a conservative
bound in view of possible unexplored parameter degeneracies.

The dependence of the neutrino mass on the cosmon is given by:
\begin{align}
\beta=-\frac{ \dd \ln m_{\rm{\nu}}(\varphi)}{\dd\varphi} < 0.
 \end{align}
In general the coupling $\beta$ can be $\varphi$-dependent. We establish in this
note that the size of the backreaction effect depends crucially on the presence
or absence of a variation of $\beta(\varphi)$. An investigation of a particle
physics motivated variation of $\beta$ \cite{Wetterich07} in reference
\cite{Ayaita14} has revealed a small backreaction effect and an overall cosmology consistent
with present observations. For a constant $ \beta$ large backreaction effects
have been observed \cite{Ayaita12a}. We adress here the question if the model
remains compatible with observations in this case as well.

A constant coupling implies for the neutrino mass:
\begin{align}
m_{\nu}(\varphi)=m_{i}e^{-\beta \varphi},
\end{align}
where an additive constant in $\varphi$ is fixed such that
$V(\varphi=0)=2.915\cdot 10^{-7}\:\rm{eV}$. The $\varphi$-dependent neutrino
mass allows for energy transfer between neutrinos and the cosmon, which is
proportional to the trace of neutrino energy momentum tensor:
\begin{align}
\nabla_{\nu}T^{\mu\nu}_{(\varphi)}&=+\beta T_{(\nu)}\dot{\varphi}\notag\\
\nabla_{\nu}T^{\mu\nu}_{(\nu)}&=-\beta T_{(\nu)}\dot{\varphi}.
\end{align}
The trace of the energy momentum tensor
$T_{(\nu)}=T^{\mu}_{\mu,(\nu)}=-\rho_{\rm \nu}+3 P_{\rm \nu}$ vanishes for
 ultra-relativistic neutrinos. The coupling between neutrinos
and the cosmon is therefore ineffective for relativistic neutrinos. The neutrino
energy-momentum tensor also sources the Klein-Gordon equation which governs the
evolution of the cosmon:
\begin{align}
\nabla_{\mu}\nabla^{\mu}\varphi-V'(\varphi)=\beta T_{\rm{(\nu)}}.\label{eq:kg}
\end{align}
We will describe neutrinos and dark matter by an N-body simulation. The
trajectories of classical neutrinos obey a modification of the
geodesic equation \cite{Ayaita12a}:
\begin{align}
\frac{\dd u^{\mu}}{\dd \tau}+\Gamma^{\mu}_{\nu \lambda}
u^{\nu}u^{\lambda}=\beta\partial^{\mu} \varphi+\beta
u^{\nu}\partial_{\nu}\varphi u^{\mu},\label{eq:eomneut}
\end{align}
where $u^{\mu}$ denotes the four-velocity and $\tau$ the proper time. The left
hand side is the usual gravitational motion, with the Christoffel symbols
$\Gamma^{\lambda}_{\mu \nu}$ determined by the metric. Throughout this work we
use the Newtonian gauge for the metric:
\begin{align}
ds^2=-\left(1+2\Psi\right)\dd t^2 +a^2\left(1-2\Phi\right) \dd \mathbf{x}^2.
\end{align}
We will work to first order in the gravitational potentials $\Phi$ and $\Psi$
and neglect there time derivatives.

The right hand side of equation \eqref{eq:eomneut} describes an additional force due to the coupling to the
cosmon. It consists of two parts.
First, a velocity dependent part $\beta u^{\nu}\partial_{\nu}\varphi u^{\mu}$ compensates changes in the mass for neutrinos moving in a varying
cosmon field so that momentum is conserved. A neutrino moving into a region with
smaller (larger) values of $\varphi$ will lose (gain) mass. To compensate the
loss (gain) of momentum it will be accelerated (decelerated).
Second, a velocity independent fifth force $\beta\partial^{\mu} \varphi$. In the
non-relativistic limit it acts as an attractive force about $2 \beta^2$ times
stronger than gravity \cite{Wintergerst10a}.

%%%%%%%%%%%%%%%%%%%%%%%%%%%%%%%%%%%%%%%
%%%%%%%%%%%%%%%%%%%%%%%%%%%%%%%%%%%%%%

\subsection{Homogeneous Evolution}
\label{sec:hom}
Let us now turn to the homogeneous limit and discuss how GNQ in its simplest
form can lead to an accelerated expansion of the universe. At early times when the neutrinos are relativistic the
evolution of the cosmon is determined by the potential. Therefore the cosmon
will evolve towards its scaling solution with the DE density decreasing with
$a^{-3}$ during matter domination.
\begin{figure}[t]
\centering
\includegraphics[width=0.49\textwidth]{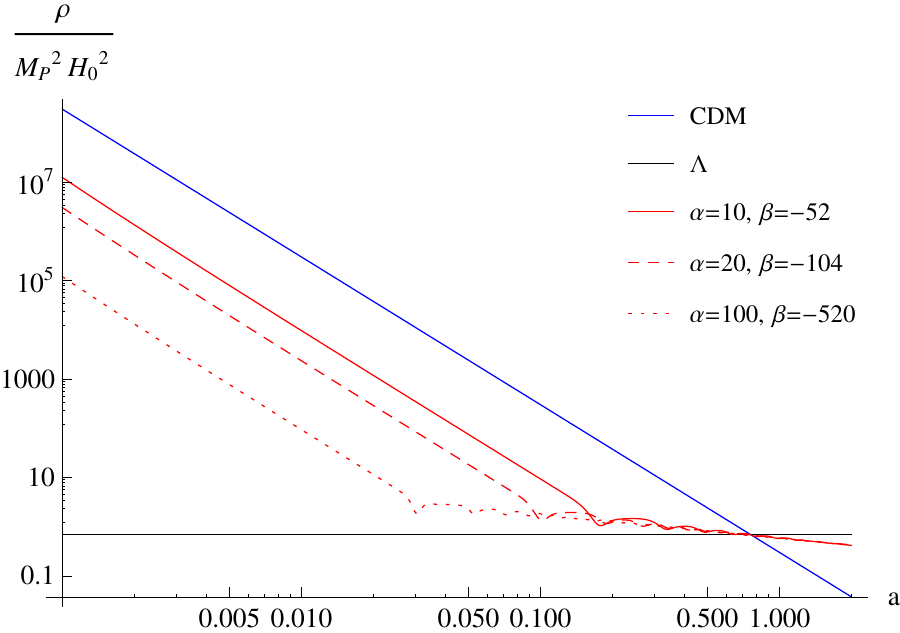}
\caption{ Energy density of the cosmon-neutrino fluid, for different
parameters $\alpha$ and $\beta$. We compare to the CDM-density and the density
of a cosmological constant $\Omega_{\Lambda}$. The parameters where chosen to match
$\Omega_{\Lambda}$ today. The stopping occurs earlier for larger $\alpha$,
with a smaller amount of early dark energy.}
\end{figure}
In view of the growing mass the neutrinos become non-relativistic rather late.
The interaction becomes important once $\alpha V(\varphi)\approx
\beta T_{(\rm{\nu})}\approx -\beta \rho_{\rm{\nu}}$. It acts as an
effective potential barrier stopping the time evolution of the energy density of
the cosmon-neutrino fluid. The constant energy density then mimics a
cosmological constant.
Since the energy density of the neutrinos is small compared to the cosmon energy density, the coupling must be rather large.

Most of the cosmological parameters as $\Omega_{\rm{DE}}=\Omega_{\rm{\varphi}}+\Omega_{\rm{\nu}}$ and
$m_{\rm{\nu}}$ are approximately independent of the individual values of $\alpha$ and
$\beta$, they only depend on there ratio. Demanding a dark energy density of
$\Omega_{\rm DE}\approx 0.7$ enforces $-\frac{\beta}{\alpha}\approx 5$
\cite{Amendola08} for a present neutrino mass $m_{\nu}=O(1\:\rm{eV})$, where
smaller neutrino masses require large $-\frac{\beta}{\alpha}$.
We note that the usual cosmological bounds on the neutrino mass from CMB and
Large Scale Structure observations \cite{Lesgourgues06,Wong11} do not apply
here, since neutrino masses have been substantially smaller in the past. In the
homogeneous limit the neutrino mass is mainly constrained by earth based
experiments.
Also the scale factor at which the neutrinos stop the cosmon evolution has only
a moderate dependence on the individual values of $\alpha$ and $\beta$. The
energy density fraction of the cosmon before stopping is given by
$\Omega_{\varphi}\propto \alpha^{-2}$ and hence becomes smaller for larger $\alpha$. The time at which the interaction with neutrinos compensates the self interaction of the cosmon becomes earlier for larger $\alpha$. The onset of dark energy is therefore earlier for larger values of $\alpha$ and $\beta$.

As we will discuss later, strong backreaction effects will alter this simple
picture. We will see in
section \ref{sec:backreact} that backreaction effects always counteract the
stopping mechanism and the cosmon will evolve again, so that it is not guaranteed that values for $\alpha$ and $\beta$ which describe a realistic cosmology in the homogeneous limit will also describe a close-to-realistic cosmology including backreaction.

Since backreaction effects can only be important after the neutrinos became
non-relativistic the homogeneous description remains valid at early times. Large
values for $\alpha$ are preferred by bounds on early dark energy. For large
$\alpha$ the stopping mechanism acts earlier, hence also the backreaction becomes important earlier. From these qualitative
considerations we already find some tension between reducing the backreaction
effects, which spoil the stopping of the cosmon evolution, and satisfying
bounds on EDE.

%%%%%%%%%%%%%%%%%%%%%%%%%%%%%%%%%%%
%%%%%%%%%%%%%%%%%%%%%%%%%%%%%%%%%%%

\section{Backreaction and Effective Equation of State}
\label{sec:backreact}
\subsection{Neutrino Lumps}
In GNQ  it is important to understand structure formation, not only in view of
using Large Scale Structure observation as a probe for our cosmological models,
especially to test DE models or ``measure'' the neutrino mass. It is
crucial to understand the formation and evolution of neutrino lumps, before
being able to judge about the viability of GNQ as a DE model. In this section we
shortly review the progress towards an understanding of the neutrino lumps, for
details we refer to previous work \cite{Mota08, Pettorino10, Ayaita12a, Ayaita12b, Brouzakis07, Wintergerst10a, Schrempp09, Nunes11, Baldi11a}. Our main focus lies on the strong backreaction
effects from non-linear perturbations in the neutrino-cosmon fluid.

\begin{figure}[t]
\centering
\includegraphics[width=0.49\textwidth]{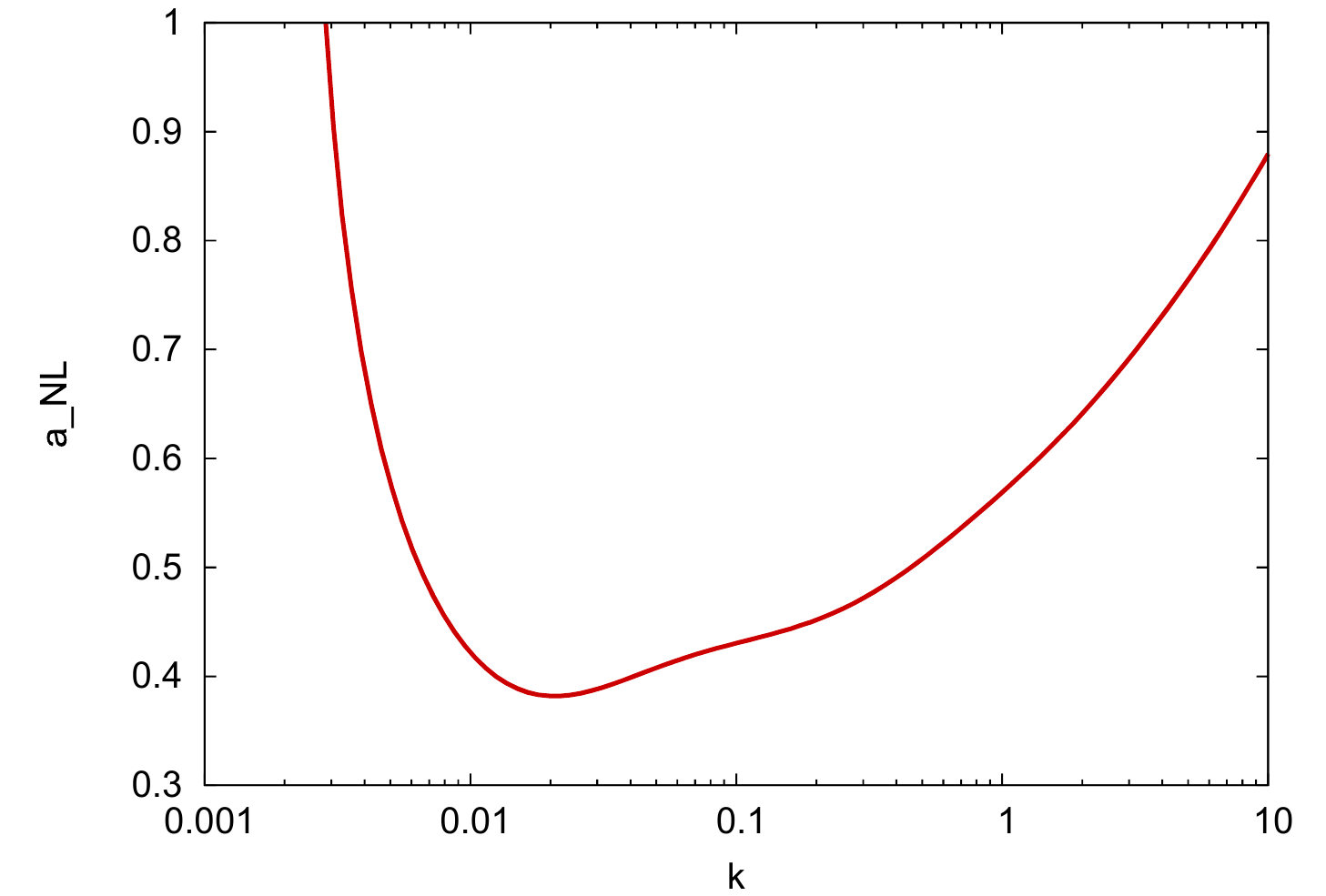}
\caption{The scale factor $a_{\rm{NL}}$ at which the dimensionless linear
neutrino power spectrum becomes unity, $\Delta(k,a_{\rm{NL}})=1$, as a function
of scale, for the parameters $\alpha=10$ and $\beta=-52$. Already at $a\sim0.40$
scales around $k\sim0.02$ are non-linear demonstrating the failure of standard
perturbative methods compare to the same figure in reference
\cite{Ayaita14}.\label{fig:knl}}
\end{figure}
The large non-linearities have there origin in the large coupling
$\beta=O(10^2)$. Therefore the additional force
between neutrinos will be about $10^3-10^4$ times larger than the
gravitational interaction between neutrinos and between neutrinos and CDM. In
turn the neutrino perturbations grow very quickly as soon as neutrinos become
non-relativistic. This implies that the fluctuations in the neutrino energy
density become non-linear even at large scales.
The scale factor $a_{\rm{NL}}$ at which this happens for a neutrino
perturbation of a given wavelength $k^{-1}$ can be estimated by the value of
$a$ at which the linear dimensionless neutrino Power Spectrum $\Delta_{\rm{\nu}}(k)=k^3 P_{\rm{\nu}}(k)/(2\pi^2)$ becomes order
unity. Looking at figure \ref{fig:knl}, we see that for the particular choice
of parameters $\alpha=10$ and $\beta=-52$ already at $a\sim 0.4$ scales around $k_{\rm{NL,\nu}}\sim 0.01\rm{\:h\:Mpc^{-1}}$ become non-linear, while today scales around $k_{\rm{NL,\nu}}\sim 0.002\rm{\:h\:Mpc^{-1}}$ are
non-linear. The exact value of
the non-linear scale of neutrino-cosmon perturbations depends on the chosen
parameters, but it is a generic finding that $k_{\rm{NL,\nu}}$  is smaller than
the corresponding wave vector for CDM perturbations, $k_{\rm{NL,C,0}}\sim
0.1\;\rm{h\;Mpc^{-1}}$. These can be traced back to instabilities
in the neutrino perturbations already present at linear order. These
instabilities are stabilized non-perturbatively by the formation of neutrino
lumps.
%%%%%%%%%%%%%%%%%%%%%%%%%%%%%%%%%%%%%%%
%%%%%%%%%%%%%%%%%%%%%%%%%%%%%%%%%%%%%%%

\subsection{Backreaction}
\label{sec:back}
\begin{figure}
\centering
\includegraphics[width=0.49\textwidth]{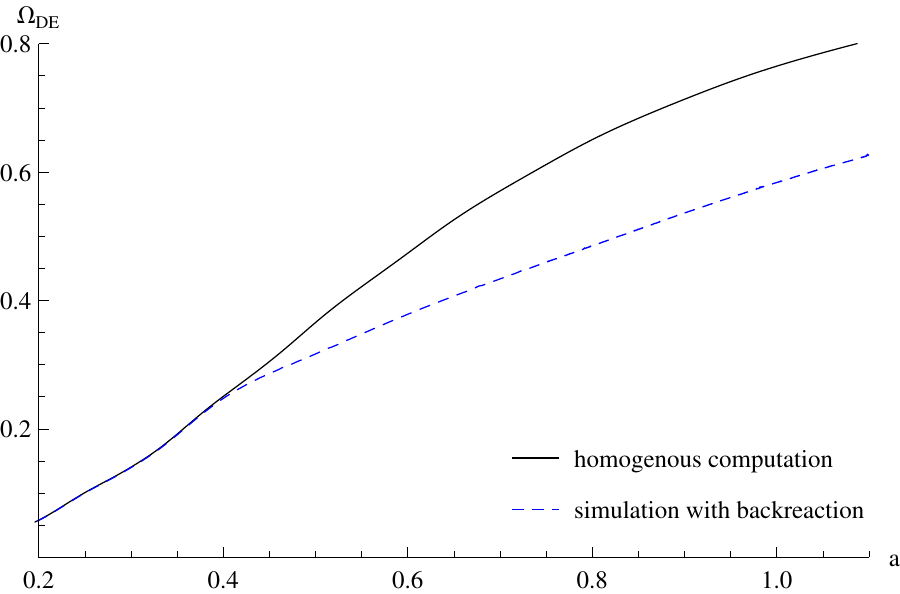}
\includegraphics[width=0.49\textwidth]{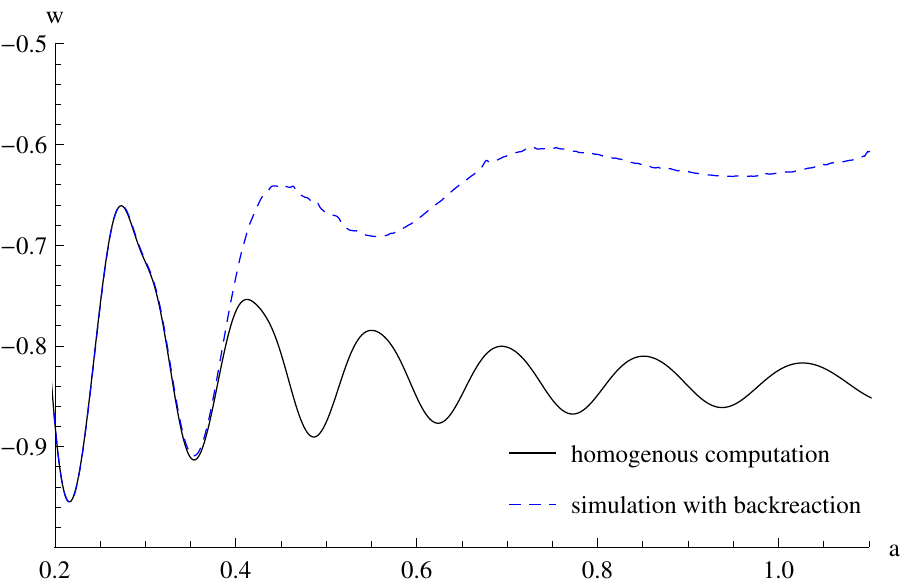}
\caption{Dark energy density fraction $\Omega_{\rm{DE}}$ (top)
and equation of state $w$ (bottom) as a function of the scale factor, for
$\alpha=10$ and $\beta=-52$, with and without backreaction.}
\label{fig:backreaction}
\end{figure}

Usually backreaction in cosmology is assumed to be negligible. In the last years
several quantitative estimates \cite{Wetterich02b, Baumann10, Adamek14} came to the
conclusion that backreaction is indeed small in the $\rm{\Lambda CDM}$-model.
In contrast, backreaction effects are crucial in GNQ. We
demonstrate this in figure \ref{fig:backreaction}, where we compare the numerical results for the clumping neutrinos with the pure background evolution
for which the effects of non-linear neutrino perturbations are neglected. We
choose the parameters $\alpha=10$ and $\beta=-52$ that have often been
employed in the literature.

We find two types of backreaction effects. First, the
Friedmann equation involves the volume averaged energy density, which we will define below.
Second, the average value of the cosmon $\bar{\varphi}$ can not be obtained by
solving the homogeneous equation of motion. The Klein-Gordon-Equation needs to
be modified to include backreaction effects from the neutrino lumps. The reason is
that the typical velocities and masses of the neutrinos do not coincide
with there counterparts of the homogeneous calculation.
While the first effect mainly affects the expansion history of the universe, the
second effect is also important for an understanding of the lump dynamics.

Let us first discuss the second effect. Due to the strong interaction most
of neutrinos are bound in the lumps. Inside gravitational bound objects the gravitational
potential has a well. Similar, inside neutrino lumps the local field value is
smaller than its average by an amount of $\delta \varphi$. The mass of a
neutrino inside a lump is therefore smaller than the mass of a free-streaming
neutrino $m(\bar{\varphi}+\delta \varphi)<m(\bar{\varphi})$ . As a
consequence most of the neutrinos have a mass
substantially smaller than the mass estimated from the homogenous calculation. Due to the
velocity dependent force the loss of mass during the formation of lumps is accompanied by an
acceleration to relativistic velocities. These two effects lead to a mismatch
between the energy momentum tensor of neutrinos from the homogeneous calculation
and its average value, as soon as the formation of lumps has started.

We account for the backreaction effects by using the volume averaged energy
momentum tensor. The Klein-Gordon equation for the average field is given
approximately by:
\begin{align}
\ddot{\bar{\varphi}}+3 H\dot{\bar{\varphi}}+\alpha
V(\bar{\varphi})=-\beta \overline{T}_{(\nu)},\label{eq:kgavT}
\end{align}
where the volume average is defined as
\begin{align}
\overline{T}_{(\nu)}=\frac{1}{V} \int d^3x\: \sqrt{g^{(3)}} T_{(\nu)} \approx
\frac{a^3}{V} \int d^3x\: \left(1-3\Phi\right) T_{(\nu)}.
\end{align}
The determinant of the spatial 3-metric up to first order in
metric perturbations is given by $\sqrt{g^{(3)}}\approx
a^3\left(1-3\Phi\right)$. The integration is to be
understood over the whole simulation box. The volume is given by $V \approx a^3 \int d^3x\: \left(1-3\Phi\right)$. Taking backreaction
effects consistently into account and evolving the volume averaged field $\bar{\varphi}$ additional modifications arise in the equation.
However, we will neglect these terms for the qualitative discussion of
backreaction in this section and postpone a more detailed discussing to
section \ref{sec:N-Body}.

The right hand side of equation \eqref{eq:kgavT} can be written as:
\begin {align}
\beta\overline{T}_{(\rm{\nu})}=\beta\left(-\bar{\rho}_{\rm{\nu}}+3\bar{P}_{\rm{\nu}}\right)=-\beta\bar{\rho}_{\rm{\nu}}\left(1-3
w_{\rm{\nu}}\right)<-\beta\bar{\rho}_{\rm{\nu}},
\end{align}
where the energy density and pressure are understood as volume averages. We use
them to define the equation of state $w_{\rm{\nu}}$. The neutrino pressure is
positive ($w_{\nu}\geq 0$) such that pressure effects lower the effective
potential barrier which stops the cosmon evolution.
As a consequence, the time at which the cosmon evolution stops is postponed
towards the future. If the evolution has already stopped the effective
reduction of the barrier can have the effect that the cosmon will evolve again.
The weaker interaction between the neutrinos and the cosmon after the formation of lumps, can also be interpreted as a lower effective coupling $\beta_{l}$, which gets renormalized by integrating out short wavelength
modes \cite{Ayaita12b}. In a qualitative sense $\beta_{l}$ can be interpreted
as the effective coupling between a fluid of neutrino lumps and the homogenous
cosmon field. The smaller value of $\beta_l$ as compared to $\beta$
is the dominant backreaction effect in our model.

We next turn to the backreaction effect for the evolution of the background
metric. One needs to replace the background density
of neutrinos and the cosmon by their volume average, such that the Friedmann
equation becomes:
\begin{align}
H^2=\bar{\rho}_{\rm{CDM}}+\bar{\rho}_{\rm{\nu}}+\bar{\rho}_{\rm{\varphi}}.
\end{align}
In the presence of lumps $\rho_{\nu}$ has contributions from the neutrino
velocities, and $\rho_{\varphi}$ involves additional gradient contributions. The
observable DE component is the combined neutrino-cosmon fluid $\rho_{\rm{DE}}$.
The neutrinos are typically subdominant but still contribute a significant
fraction $\frac{\bar{\rho}_{\rm{\nu}}}{\rho_{\rm{DE}}}\sim 0.1$. With an
equation of state $w_{\rm{\nu}}\sim 0.1$ the neutrinos lift the dark energy equation of state away from $w\approx-1$ to some higher value.

The volume average of the cosmon energy density is given by:
\begin{align}
\bar{\rho}_{\varphi}=\frac{1}{2}\overline{\dot{\varphi}^2}+\frac{1}{2
a^2}\overline{(1+2
\Phi)\left(\partial_{i}\varphi\right)
\left(\partial_{j} \varphi\right)\delta^{ij}}+\overline{V(\varphi)},
\end{align}
where we only keep metric perturbations up to first order, neglected their time
derivatives and use that the volume average of the gravitational
potentials vanishes $\bar{\Phi}=\bar{\Psi}=0$.
Also assuming that time derivatives of the cosmon perturbation
$\delta \varphi$ are small allows us to approximate
$\overline{\dot{\varphi}^2}\approx\dot{\bar{\varphi}}^2$.
Using the quasi static approximation is justified although the individual neutrino velocities are large. For the quasi static approximation to hold it is sufficient
that the energy-momentum tensor for all neutrinos does not evolve fast, so that there
are no fast varying sources for the cosmon.
A non-zero $\delta \dot{\varphi}$  results in a positive contribution to the
pressure, making it even harder to achieve an almost constant energy density for
the cosmon-neutrino fluid.

Without the gradient term one has the usual competition between
potential and kinetic energy. The potential energy should be dominant
in order to have an accelerated expansion. The averaged potential
energy $\overline{V(\varphi)}$ differs from the potential energy
$V(\bar{\varphi})$ of the averaged field $\bar{\varphi}$ only by a few percent,
such that no major backreaction effect arises from this source.
In contrast, the gradient term can be almost as large as the the potential
energy. From the expression for the pressure
\begin{align}
\bar{P}_{\varphi}\approx\frac{1}{2}\dot{\overline{\varphi}}^2-\frac{1}{6
a^2}\overline{(1+2\Phi)\left(\partial_{i}\varphi\right)
\left(\partial_{j} \varphi\right)\delta^{ij}}-\overline{V(\varphi)},
\end{align}
we see that a gradient term dominated equation of state would be
$w_{\nu}=-\frac{1}{3}$. We emphasize that all backreaction effects
individually lead to an evolving energy density of neutrino-cosmon fluid and
typically push $w$ away from $-1$.

For models with constant $\beta$ the lumps have the tendency to stabilize and
to remain present once formed. The neutrino-cosmon fluid can be understood as an
effective fluid of nearly virialized neutrino lumps with parameters differing
from the microscopic ones \cite{Ayaita12b}. The observable DE is then the sum of
a neutrino lump fluid and a homogenous background field. For virialized lumps the
pressure between relativistic neutrinos and cosmon gradients is expected to
cancel \cite{Ayaita12b}. Therefore the equation of state of the
lump fluid is close to zero, similar to the fluid of non-relativistic
neutrinos. The backreaction effect that remains
even in this limit is the reduced effective coupling $\beta_{l}$ between
neutrino lumps and the cosmon background field. Due to the not completely virialized lumps the pressure contribution from the neutrinos
and the cosmon gradients do not cancel exactly, adding a small but relevant
additional backreaction effect. This is different to gravitationally bound
objects, for which a non-renormalization theorem states that small virialized
objects decouple completely from the background evolution and there is no backreaction effect from small virialized objects at all \cite{Baumann10}.

%%%%%%%%%%%%%%%%%%%%%%%%%%%%%%%%%%%
%%%%%%%%%%%%%%%%%%%%%%%%%%%%%%%%%%%

\section{N-Body Simulation}
\label{sec:N-Body}
The highly non-linear nature of the neutrino lumps makes there description
non-amenable to standard perturbative techniques. Instead we use a N-body simulation
specially designed for GNQ. The N-Body simulation solves the background and the inhomogeneities
simultaneously and therefore allows us to study the backreaction effect of lumps
on the homogeneous background evolution. Concept and many details of the
simulation were already described in \cite{Ayaita12a,Ayaita14}, we focus here on the equation of motion for the average cosmon field $\bar{\varphi}$ and its perturbation 
$\delta \varphi$.

In our simulation we follow the usual motion of non-relativistic CDM particles
and there clustering due to gravity. In contrast to the standard picture of
structure formation the two gravitational potentials differ, $\Phi\neq \Psi$,
because of the anisotropic stress from the neutrinos. This is accounted for by
solving the Poisson equation for $\Phi-\Psi$, which yields $\Phi$ ones the
Newtonian potential $\Psi$ is known. The Poisson equation for $\Psi$ is sourced
by the energy density of CDM, neutrinos and to a small part
by the one of the cosmon perturbations.

The neutrinos are evolved using equation \eqref{eq:eomneut}. The cosmon
evolution is governed by the Klein-Gordon equation \eqref{eq:kg}. We split the
cosmon into the volume average $\bar{\varphi}=\frac{1}{V}\int d^3x \:
\sqrt{g^{(3)}} \varphi$ and a perturbation $\delta \varphi=
\varphi-\bar{\varphi}$. Neglecting time derivatives of the gravitational
potentials, time derivatives commute with the process of averaging 
$\dot{\bar{\varphi}}\approx\bar{\dot{\varphi}}$. The averaged equation
\eqref{eq:kg} is:
\begin{align}
\ddot{\bar{\varphi}}+3 H
\dot{\bar{\varphi}}+\alpha\overline{(1+2\Psi)V(\varphi)}\notag\\
=-\beta \overline{(1+2\Psi)T_{(\nu)}}+a^{-2}\overline{\delta^{ij}\left(\partial_{j}\Psi\right)\left(\partial_{i}\delta\varphi\right)},\label{eq:kgav}
\end{align}
where we expanded up to first order in metric perturbations. Equation \eqref{eq:kgav}
is the full version of equation \eqref{eq:kgavT}. As already discussed in section
\ref{sec:backreact} the most important difference as compared to a naive
homogeneous calculation is the use of the actual average of the neutrino
momentum tensor. Including the gravitational potential in the average gives only
a minor correction. Also the averaged potential term agrees up to a few percent
with the homogeneous estimate. The gradient terms is roughly one order of
magnitude smaller than the potential term and therefore only subdominant.
Nevertheless, our numerical code includes all these effects.

By subtracting equation \eqref{eq:kgav} from the the Klein-Gordon equation
\eqref{eq:kg} we find the equation for the perturbation:
\begin{align}
\delta
\ddot{\varphi}+3H\delta\dot{\varphi}-a^{-2}\delta^{ij}\partial_{i}\partial_{j}\delta\varphi(1+2\Phi)\notag\\
-a^{-2}\delta^{ij}\left(\partial_{j}(\Psi-\Phi)\right)\left(\partial_{i}\varphi\right)+a^{-2}\overline{\delta^{ij}\left(\partial_{j}\Psi\right)\left(\partial_{i}\varphi\right)}\notag\\
+\alpha\left((1+2\Psi)V(\varphi)-\overline{(1+2\Psi)V(\varphi)}\right)\notag\\
=-\beta
\left((1+2\Psi)T_{(\nu)}-\overline{(1+2\Psi)T_{(\nu)}}\right).\label{eq:kgpert}
\end{align}
This equation is a non-linear wave equation, which is, due to the averaging,
non-local in position space. To be able to solve this equation we need to make
some approximations. We employ a quasi static approximation for the cosmon
perturbation for which we neglect the second order time derivative $\delta
\ddot{\varphi}$. Simply neglecting all time derivatives is not a consistent
approximation. Doing so the resulting equation does not ensure that the
perturbation has a vanishing mean $\overline{\delta \varphi}=0$. This can be
seen by averaging equation \eqref{eq:kgpert}. Taking into account the $\Phi$
dependence in the volume element  and only keeping terms to first order in
the metric perturbations all terms except the time derivatives cancel:
\begin{align}
\ddot{ \overline{\delta\varphi}}+3H\dot{\overline{\delta\varphi}}=0.
\end{align}
This relation ensures that the if the average vanishes initially it will vanish
at all times. This is still true if we neglect the second time derivative while
keeping the first one. This approximation is consistent with the
approximation for kinetic term of the average energy density and pressure
\begin{align}
\overline{\dot{\varphi}^2}=\dot{\bar{\varphi}}^2+\overline{\delta\dot{\varphi}^2},
\end{align}
where we neglected the $\overline{\delta\dot{\varphi}^2}$-term. So we neglected
all terms which are second order in the time derivatives of the cosmon
perturbations these terms are smaller than those with only one time
derivative.

If one instead neglects the second derivative with respect to conformal time the Hubble damping changes $3H\rightarrow 2H$, we compared both possibilities and found only a small difference. We interpret this a sign that the quasi static approximation is justified.\\
To solve the equation for $\delta \varphi$ we use a Newton-Gauß-Seidel (NGS)
multigrid relaxation method, already applied to the varying coupling model
\cite{Ayaita14} and originally developed for modified gravity \cite{Puchwein13}.
The quasi static approximation is crucial for applying the NGS method, which is
not applicable to wave like equations, but can be applied to diffusion like
equations \cite{vanLent06}. The idea of the NGS solver is to rewrite the
equation to be solved into a functional form:
\begin{align}
\mathcal{L}[\delta \varphi]=D\delta \varphi-F[\delta \varphi]=0,
\end{align}
with some differential operator $D$ and a non-linear functional $F$. The root of $\mathcal{L}[\delta \varphi]=0$ can be obtained by a newton-like iterative procedure:
\begin{align}
\delta\varphi ^{(n+1)}=\delta\varphi ^{(n)}-\mathcal{L}[\delta\varphi
^{(n)}]\left(\frac{\partial \mathcal{L}[\delta\varphi ^{(n)}]}{\partial
\delta\varphi ^{(n)}}\right)^{-1},
\end{align}
the derivative is taken at each point individually, the coupling between
different points, induced by the derivatives, is taken into account solely by
the iterative procedure. The derivative of the differential operator
$\frac{\partial D \delta\varphi}{\partial \delta\varphi}$ is defined by the
discretisation rule used in the simulation. We define the gradient and
the laplacian by relating a grid point to its neighbors in
$j$-direction by a Taylor expansion: $\delta \varphi(x_i\pm\Delta x
\delta_{ij})=\delta \varphi(x_i)\pm\partial_j\delta \varphi(x_i)\Delta x+\frac{1}{2}
\partial^2_j\delta\varphi(x_i)\Delta x^2 +\ldots$, with $\Delta x$ the spacing
between two grid points. The laplacian is then approximated by a seven-point
stencil and the derivative is $-6/\Delta x^2$. The derivative of the
gradient vanishes.

In principle this method can be applied even in the presence of the non-local
terms present in equation \eqref{eq:kgpert}.In practice this not
possible because calculating the non-local terms involves an integration over the full
simulation box in each iteration step.
We account for these terms iteratively. The difference between the values of the average terms of two time steps is small. So we use at a given time step the average terms of the proceeding time step as first approximation and apply the NGS solver a few times to correct for the difference.

%%%%%%%%%%%%%%%%%%%%%%%%%%%%%%%%%%%
%%%%%%%%%%%%%%%%%%%%%%%%%%%%%%%%%%%

\section{Results and Discussion}
\label{sec:Results}
Using the N-Body simulation described in section \ref{sec:N-Body} we perform a
parameter scan and search for parameters describing a realistic universe with accelerated expansion. For the details on the formation of lumps and there characteristics we refer to previous work
\cite{Ayaita12a,Ayaita12b}. We use a simulation box with a comoving volume of
$V=(600\:h^{-1}Mpc)^3$, which we divide into $N_{\rm{c}}=128$ cells. The number
of effective CDM particles $N_{\rm{C}}$ and neutrino particles $N_{\rm{\nu}}$ is
chosen to be equal to the number of cells $N_{\rm{c}}=N_{\rm{C}}=N_{\rm{\nu}}$. The initial power spectrum has a spectral index of $n_{s}=0.96$ and an amplitude of $A_{s}=2.3\cdot 10 ^{-9}$ at the pivot scale
$k_{\rm{pivot}}=0.05\:\rm{Mpc}^{-1}$. We start our simulation with the CDM
particles only at $a_{\rm{ini, C}}=0.02$ and add the neutrinos at a later time,
after they became non-relativistic.

\begin{figure}[t]
\centering
\includegraphics[width=0.49\textwidth]{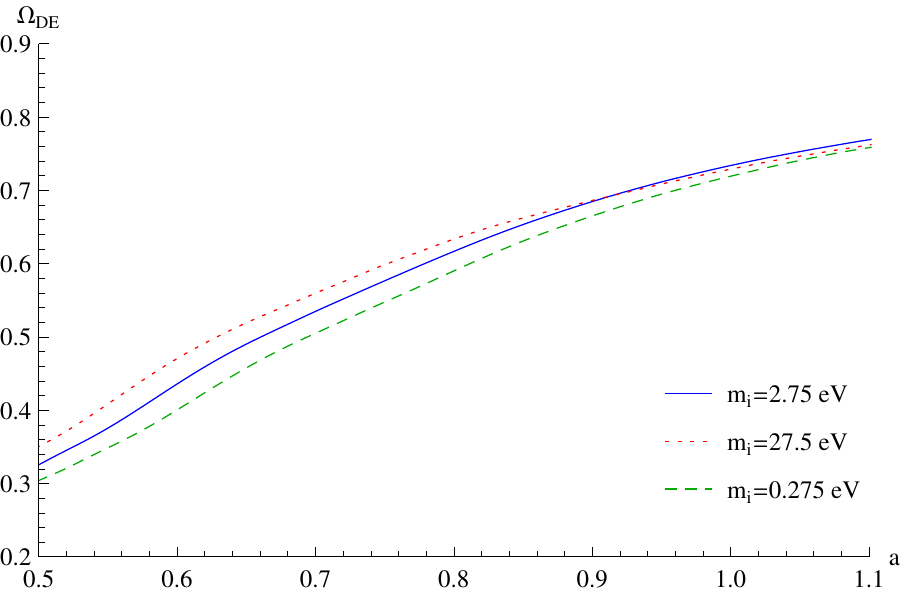}
\includegraphics[width=0.49\textwidth]{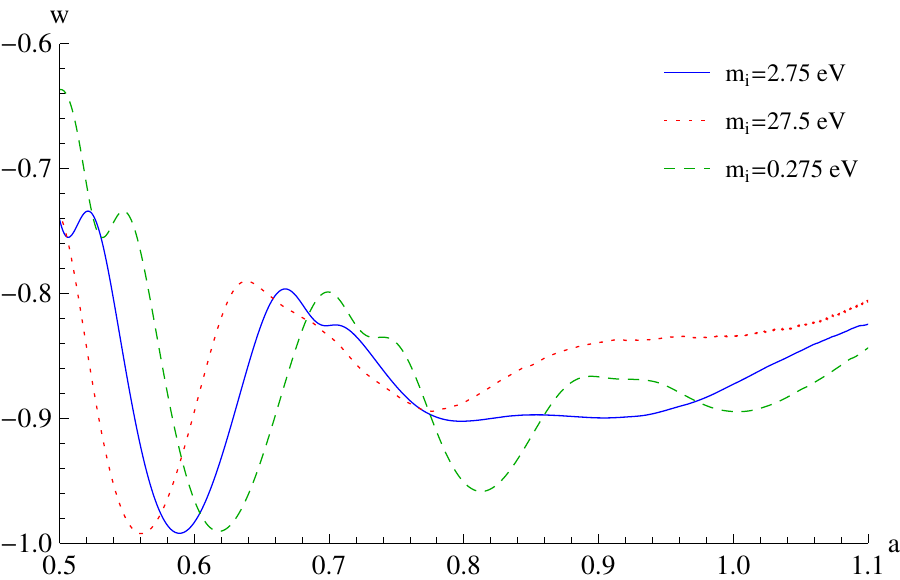} 
\caption{Energy density fraction of the cosmon-neutrino fluid
$\Omega_{\rm DE}$ and effective equation of state $w$, for different
mass parameters $m_i$, with $\alpha=5$ and $\beta=-78$. Even for mass parameters
different by a factor of 100 the equation of state varies at
maximum about $10\%$, indicating that the value of $m_{i}$ plays only a minor
role.\label{fig:diffm}}
\end{figure}
In view of the strong backreaction effects it is no longer clear that the
stopping power of neutrinos for the time evolution of the cosmon is sufficient
in order to account for a large present fraction of dark energy and
an acceleration of the expansion similar to a cosmological constant. If so,
the parameter range where this happens may be rather different from the one where the background
evolution neglects the effect of neutrino structures.

Our model has three parameters relevant for this investigation, namely
$\alpha$, related to the amount of EDE, the neutrino cosmon
coupling $\beta$ and  $m_{i}$, related to the size of the neutrino
mass. We have performed a parameter scan in order to search for a
parameter range consistent with observations. For this purpose we vary the
parameters $\alpha$ and $\beta$ individually while fixing the mass parameter to
$m_{i}=1\:eV$. Figure \ref{fig:diffm} shows that changing the mass parameter by a factor of 10 effects the effective equation of state and the
energy density by no more than $10\%$.

\begin{figure}
\centering
\includegraphics[width=0.49\textwidth]{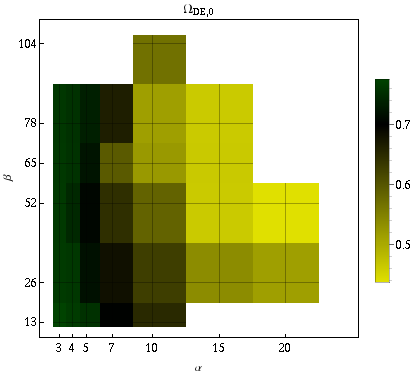}
\includegraphics[width=0.49\textwidth]{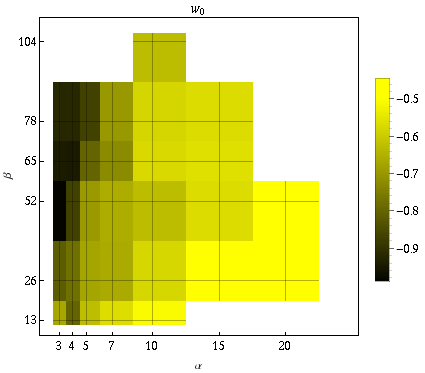}
\caption{Present energy density $\Omega_{\rm{DE},0}$ and equation of state
$w_0$ of the cosmon-neutrino fluid. Realistic values ($w_{0}\approx-1$,
$\Omega_{\rm{DE},0}\approx0.7$ ) are found for small values of $\alpha$.
It is hard to get both values ``correct'' simultaneously, for sufficiently
large $\alpha$.
\label{fig:parameter}}
\end{figure}

A realistic DE model must certainly assume the benchmark
values for the present DE density $\Omega_{\rm{DE},0}\approx0.7$ and
the present equation of state $w_{0}\approx-1$. In figure \ref{fig:parameter} we
show the values of $\Omega_{\rm{DE},0}$ and $w_{0}$ for a grid in
the parameter space for $\alpha$ and $\beta$. Sufficient acceleration typically
requires rather small values $\alpha\lesssim5$. A band with an acceptable
fraction of present DE is typically found in the range
$5\lesssim\alpha\lesssim 10$, showing some tension already at this stage.

The parameter range yielding
an accelerated expansion ($\alpha \lesssim 5$) is problematic also in view of the
bounds on EDE, which require $\alpha \gtrsim 10$. In the parameter range where
one finds $w_{0}<-0.9$ some tension persists if one tries to get both the
equation of state and the energy density compatible with observations. For $\alpha=3$ and $\alpha=4$ we indeed find $w_0\lesssim -0.9$ but the energy density exceeds with $\Omega_{\rm{DE}}\approx0.75$ the benchmark value of $\Omega_{\rm{DE},0}\approx0.7$. On the other hand
for $\alpha=5$ one has $\Omega_{\rm{DE}}\approx 0.7$, but the equation of
state is $w_0\approx -0.7$. Although we could not find parameters for which
$w_0$ and $\Omega_{\rm{DE},0}$ match the benchmark values precisely, our results
are not too far from those values either. It might be that varying also the mass
parameter $m_{i}$ could bring them into agreement with observations.

\begin{figure}
\centering
\includegraphics[width=0.49\textwidth]{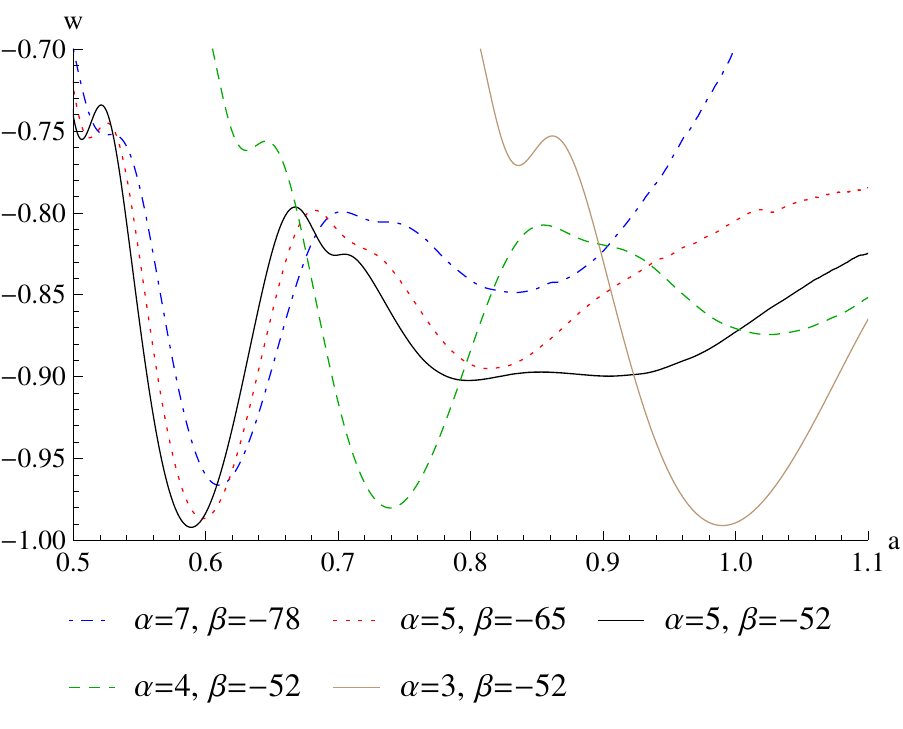}
\caption{Equation of state as a function of the scale factor. The
model parameters are chosen such that $w$ and $\Omega_{\rm{DE},0}$ are near the
benchmark values. Values $w_0\lesssim -0.9$ are only reached before backreaction
effects become important. Thus $w\approx-0.99$ for $\alpha=5$ and
$\beta=-52$ is not accompanied by large negative $w$ at redshifts relevant for
supernova observations.}
\label{fig:wevolve}
\end{figure}

The equation of state is not constant in time, it can even
possess oscillating features, see figure \ref{fig:diffm}. It may happen that
the present time coincides with a minimum (maximum) of $w$ during an oscillation. In this
case the cosmic evolution is actually better described by an average value
somewhat larger (smaller) than $w_0$. The time evolution of the equation of
state is shown in figure \ref{fig:wevolve} for a range of parameters $\alpha$
and $\beta$ in the region not too far from the benchmark values. One typically
observes a first stop of the scalar field ($w\approx-1$). Due to backreaction
this is followed by a slow decrease of the dark energy, typically with
$-0.9\lesssim w \lesssim-0.8$.

Only looking at the energy density and the equation of state today is not sufficient.
In the parameter range acceptable for the benchmark the neutrinos become
non-relativistic late. Consequently the cosmon evolution stops late. This is
visible in figure \ref {fig:wevolve}, where the first pronounced minimum in $w$
precisely corresponds to the time when the increase of $\varphi$ is first
stopped and the oscillations set in. Supernova observations probe the expansion history
up to redshifts higher than $z\approx1$ and prefer an almost constant Dark
Energy \cite{Betoule14}. We find that for close-to-realistic models the equation of state reaches values around $-1$ only for scale factors $a\gtrsim 0.6$, which is difficult to get
into agreement with $w\approx-1$ from $a\lesssim 0.5$ until today.

Figure \ref{fig:wevolve} shows the generic evolution of the equation of
state: It drops down after the neutrinos became non-relativistic followed by a
few damped oscillations. In the homogeneous evolution these oscillations are
damped away quickly and the equation of state assumes an almost constant value
rather close to $w=-1$.
In fact the equation of state grows again due to the backreaction and typically
reaches values $w\approx-0.8$.
An equation of state of $w_0\lesssim-0.9$ is only reached before or shortly
after backreaction becomes important. This simply means that lumps had not
enough time to grow large enough for being able to induce significant backreaction effects.

From these results we conclude that GNQ with a constant coupling $\beta$
is probably not a viable DE model. Realistic values for $w_0$ and
$\Omega_{\rm{DE},0}$ seem only possible if the cosmon evolution is stopped late,
so that backreaction effects have no time to become important. Stopping the
cosmon evolution late is in some tension with supernova data and
involves a large amount of EDE, probably not consistent with observations.

%%%%%%%%%%%%%%%%%%%%%%%%%%%%%%%%%%%
%%%%%%%%%%%%%%%%%%%%%%%%%%%%%%%%%%%

\section{Conclusion}
\label{sec:conclusion}
We have performed a numerical analysis of Growing Neutrino Quintessence with a
constant cosmon-neutrino coupling $\beta$. Due to strong backreaction effects
from the formation of large neutrino lumps these models have difficulties to be
compatible with the observed properties of dark energy.

A specific choice for the model
parameters $\alpha$, $\beta$ and $m_i$, which appears to be compatible with
observations at the homogenous level, is typically no longer viable if
backreaction is included. Our parameter scan reveals regions for which the
backreaction effects are small enough to allow a slowly evolving cosmon and
consequently an almost constant DE density. However, this is only possible if
the neutrino lumps form late so that backreaction effects are still small today.
In this case an accelerated expansion is only possible for scale factors
$a\gtrsim 0.6$, in tension with an almost constant equation of state for scale factors
$a\lesssim0.5$, as preferred by supernova data.
Furthermore, the parameter region for which the equation of state is close to $-1$ and the DE density is not too far
from $0.7$, requires $\alpha\lesssim 5$. This contradicts constraints on early
dark energy for which $\alpha\gtrsim 10$ is necessary. We conclude that growing
neutrino quintessence with a constant coupling $\beta$ is probably not a viable DE model.

These results for a constant coupling should be contrasted with
models where $\beta$ increases with $\varphi$. For this second class of models
the backreaction effect is found to be small since the neutrino lumps form and
disrupt periodically \cite{Ayaita14}.
At the present stage this second class of models seems compatible with
observations. In certain parameter ranges it may even be hard to detect a
difference from the $\Lambda$CDM models and its variants.

These two classes of models may be seen as particular points in a larger
class of models where $\beta$ is allowed to vary with $\varphi$. Having
established points that are viable with only rather small deviations from
$\Lambda$CDM, as well as other points where the deviations are so strong that
the model is no longer acceptable, we can conclude by continuity that in between
there will be models which are still compatible with observations today, but
also offer highly interesting prospects of finding deviations from
$\Lambda$CDM. Finding large neutrino lumps, thereby
observing the cosmic neutrinos directly, would be a direct hint for GNQ.
Even for models with small neutrino perturbations we
expect observable deviations from the $\Lambda$CDM model, due to the different
evolution of the neutrino sector. First, the transition of relativistic to
non-relativistic standard massive neutrinos is imprinted in the CMB fluctuations
as well as in the matter distribution, with a specific scale dependence \cite{Lesgourgues06,Wong11}. The signal differs for constant or time-varying neutrino masses. Second, free-streaming standard massive neutrinos
attenuate the growth of matter perturbations on small scales and therefore add
an additional scale dependent effect to the matter distribution. Observing these
scale dependent effects as predicted for standard neutrinos with a constant
mass would be a strong argument for the $\Lambda$CDM model and against GNQ.

The result for models with constant $\beta$ presented in this note as well
the results on the varying $\beta$ model presented in \cite{Ayaita14} suggest that only those
models are viable in which the small scale non-linear neutrino perturbations
have only a moderate effect on the large scale dynamics. Nevertheless, the
neutrino lumps can have a observable effects on larger scales. One possibility
to account for these effects is to construct an effective fluid for the long
wavelength perturbations by averaging over small scales non-linearities as proposed
in reference \cite{Baumann10}. A similar route has already been taken in
\cite{Ayaita12b} to describe the large scale dynamics of virialized neutrino lumps in the
constant $\beta$ model by means of an effective lump fluid.
These ideas where already successfully applied to the mildly non-linear regime of
structure formation in the form of the Effective Field Theory of Large Scale Structure \cite{Carrasco12,Hertzberg12,Pajer13, Mercolli13,Carrasco13,Carrol13}, see also \cite{Rigopoulos14}.
Adopting these ideas to GNQ we hope that it will become possible to
study the dynamics of perturbations in GNQ on large
scales qualitatively. It might even become possible to study some
effects of lumps on the CMB, without running time consuming simulations.
%%%%%%%%%%%%%%%%%%%%%%%%%%%%%%%%%%%
%%%%%%%%%%%%%%%%%%%%%%%%%%%%%%%%%%%

\begin{acknowledgments}
	The authors are thankful to Youness Ayaita, Valeria Pettorino and Santiago Casas 
	for numerous inspiring discussions and ideas. They would also like
	to thank Ewald Puchwein for providing his NGS code and
	David Mota for providing his numerical implementation of linear perturbation
	theory in GNQ. FF acknowledges support from the IMPRS-PTFS and the DFG through the TRR33 project ``The Dark Universe''.
\end{acknowledgments}

% Bibliography

\bibliography{constbeta}

\end{document}